\documentclass[aps,pra,twocolumn,superscriptaddress,showpacs,10pt]{revtex4-1}

\usepackage{graphicx}
\usepackage{epstopdf}
\usepackage{amsmath}
\usepackage{verbatim} 
\usepackage{color}   
\usepackage{subfigure}
\usepackage{amssymb}
\usepackage{amsmath}
\usepackage{mathrsfs}
\usepackage{amsfonts}
\usepackage{braket}
\usepackage{appendix}
\DeclareMathAlphabet{\mathpzc}{OT1}{pzc}{m}{it}

\begin{document}


\title{Universal three-body recombination and Efimov resonances\\ in an ultracold Li-Cs mixture}

\author{J. Ulmanis}
\author{S. H\"afner}
\author{R. Pires}
\affiliation{Physikalisches Institut, Ruprecht-Karls-Universit\"{a}t Heidelberg, Im Neuenheimer Feld 226, 69120 Heidelberg, Germany}

\author{F. Werner}
\affiliation{Laboratoire Kastler Brossel, ENS-PSL, UPMC-Sorbonne Universit\'{e},
Coll\`{e}ge de France, CNRS, 24 rue Lhomond, 75231 Paris Cedex 05, France}

\author{D.~S.~Petrov}
\email[]{dmitry.petrov@u-psud.fr}
\affiliation{LPTMS, CNRS, Univ. Paris Sud, Universit\'e Paris-Saclay, 91405 Orsay, France}

\author{E. D. Kuhnle}
\affiliation{Physikalisches Institut, Ruprecht-Karls-Universit\"{a}t Heidelberg, Im Neuenheimer Feld 226, 69120 Heidelberg, Germany}

\author{M. Weidem\"uller}
\email[]{weidemueller@uni-heidelberg.de}
\affiliation{Physikalisches Institut, Ruprecht-Karls-Universit\"{a}t Heidelberg, Im Neuenheimer Feld 226, 69120 Heidelberg, Germany}
\affiliation{Synergetic Innovation Center of Quantum Information and Quantum Physics, and Hefei National Laboratory for Physical Sciences at Microscale, University of Science and Technology of China, Hefei, Anhui 230026, People's Republic of China}

\date{\today}

\begin{abstract}
We study Efimov resonances via three-body loss in an ultracold two-component gas of fermionic $^6$Li and bosonic $^{133}$Cs atoms close to a Feshbach resonance at 843~G, extending results reported previously [Pires \textit{et al.}, Phys. Rev. Lett. 112, 250404 (2014)] to temperatures around 120~nK. The experimental scheme for reaching lower temperatures is based upon compensating the gravity-induced spatial separation of the mass-imbalanced gases with bichromatic optical dipole traps. We observe the first and second excited Li-Cs-Cs Efimov resonance in the magnetic field dependence of the three-body event rate constant, in good agreement with the universal zero-range theory at finite temperature [Petrov and Werner, Phys. Rev. A 92, 022704 (2015)]. Deviations are found for the Efimov ground state, and the inelasticity parameter $\eta$ is found to be significantly larger than those for single-species systems.

\end{abstract}


\maketitle

\section{Introduction}
\label{intro}

In the 1970s Vitaly Efimov predicted the emergence of a discrete scaling symmetry for a system of three particles with pairwise resonant interactions  \cite{Efimov1971,Braaten2006}. Important consequences of these findings are the self-similarity of the system upon multiplying the two-body scattering length $a$ by a scaling factor $\lambda$ and the associated log-periodic $a$-dependence of three-body observables. In ultracold gases, where $a$ can be tuned to any value by using Feshbach resonances \cite{Ferlaino2011,Wang2013,Chin2010}, the Efimov effect can be observed by measuring the log-periodic $a$-dependence of the three-body event rate constant. Resonant enhancements of this function are expected for $a=a_-^{(n)}=\lambda^n a_-^{(0)}$ where the $n$-th Efimov trimer passes through the three-atom dissociation threshold.

In a number of experiments single Efimov resonances ($ a<0 $) have been observed in three-body losses, where the value $a_{-}^{(0)}$ was associated with the ground state of the Efimov series~\cite{Kraemer2006,  Gross2009, Pollack2009, Zaccanti2009, Gross2010,Berninger2011, Wild2012, Roy2013,Barontini2009,Bloom2013,Dyke2013}. However, the large scaling factor for equal mass systems of $\lambda = 22.7$ demands a high level of control of the magnetic field and temperature, whereupon the first excited three-body resonance was seen only recently at extremely low temperatures in a shallow trap for $^{133}$Cs \cite{Huang2014}. At positive $a$, manifestations of Efimov physics were observed as atom-dimer resonances~\cite{Pollack2009,Dyke2013,Zaccanti2009,Knoop2009,Machtey2012,Machtey2012a,Zenesini2014},  radio-frequency association of trimers \cite{Machtey2012a}, and interference minima in three-atom losses~\cite{Pollack2009,Dyke2013,Zaccanti2009,Gross2009,Gross2010}; in particular, two such minima separated by a factor $\approx \lambda$ were reported \cite{Zaccanti2009}. Efimov physics also occurs in three-component Fermi gases, as revealed by a series of experiments with $^6$Li \cite{Ottenstein2008,Huckans2009,Williams2009, Lompe2010, Nakajima2010,Naidon2011}.
Finally, the $^4{\rm He}_3$ Efimov trimer was detected recently in a molecular beam experiment~\cite{Kunitski2015}.

To measure a series of Efimov resonances ultracold gases consisting of two different species with unequal masses are experimentally more favorable, as the scaling factor $\lambda$ can be much smaller than 22.7. In K-Rb systems the Efimov ground state was investigated but a scaling factor of about 131 reduces the likelihood of observing the excited Efimov state \cite{Braaten2006,Barontini2009,Bloom2013,Hu2014}. For Li-Rb, one Efimov resonance was recently reported~\cite{Maier2015} and the observation of excited state resonances might be feasible in near future. 
In a $^{6}$Li-$^{133}$Cs system the predicted scaling factor of 4.9 is much smaller than for equal-mass systems \cite{Efimov1973} and, indeed, a series of Efimov resonances has been experimentally observed \cite{Pires2014, Tung2014,Ulmanis2015}. The visibility of higher excited Efimov resonances is limited by finite-temperature effects; when $|a|$ becomes much larger than the thermal de Broglie wavelength, the peaks in the loss rate become unresolvable and the loss rate saturates to a constant value proportional to the inverse temperature squared~\cite{DIncao2004,Kraemer2006,Rem2013,Fletcher2013,Eismann2015}. Thus, for observing the Efimov universality over a large span of length scales it is essential to prepare the coldest possible sample.

The simplest Efimov scenario (exact log-periodicity of observables) requires one to proportionally change {\it all} relevant length scales, namely the inter- and intraspecies scattering lengths as well as de Broglie wavelengths of the gas. Concurrently, they must be much larger than the short-range length scales associated with the interaction ranges (in our case these are van der Waals lengths).
In addition, the three-body parameter, which fixes the phase of the three-body wave function at short distances, is required to stay constant. It is clear that in practice each of these conditions is to some extent violated. Nevertheless, the non-proportionality of the scattering lengths and the de Broglie wavelengths, if they remain large compared to the interaction ranges, does not drive the system out of the Efimovian universality regime. These effects can be taken into account by the finite-temperature zero-range theory~\cite{Petrov2015}, for which the three-body parameter is an external parameter. Although in this case the exact log-periodicity of three-body observables is not expected, the Efimov scenario can still be checked by analyzing how well a zero-range theory curve fits experimental data. This procedure has recently been employed for characterizing low-lying Efimov resonances and retrieving three-body parameters from finite-temperature loss data in the case of identical bosons~\cite{Huang2015a,Huang2014} and equal-mass three-component fermions~\cite{Huang2014a}.

In this work, we investigate the magnetic field dependence of the three-body loss in an ultracold Li-Cs mixture close to a Feshbach resonance, and study the influence of temperature on the Efimov scenario. In our previous setup \cite{Pires2014} a crossed optical dipole trap at a single wavelength was used, and the mixing of the two gases at low temperature was limited when the atomic clouds started to separate due to a strong influence of gravity. Using the approach of species-specific traps, which has been discussed, e.g., for optical lattices \cite{LeBlanc2007}, we present a setup with two almost independent optical traps for Li and Cs atoms that allows us to control their spatial overlap. This enables a further reduction of the trap laser intensity and thus a factor of four lower temperatures as compared to the previous measurements at 450~nK \cite{Pires2014}. For a mixture prepared at 120~nK we measure the recombinational loss around 843~G and observe the first and second excited Efimov resonances. The data at both temperatures are compared to a universal zero-range model for the three-body event rate constant in heteronuclear mixtures based on the \textit{S}-matrix formalism assuming constant three-body and inelasticity parameters \cite{Petrov2015}. The model explicitly takes into account finite temperatures and the Cs-Cs background scattering length.

In Sec.~\ref{sec:1} the experimental setup for reaching temperatures around 100~nK in a Li-Cs mixture is elucidated. In Sec.~\ref{sec:2} we present new measurements of the three-body event rate constant at 120~nK and analyze the observed Efimov resonances with the zero-range model.

\section{Experimental procedure and compensation of spatial separation}
\label{sec:1}

\begin{figure*}
	\centering
	\includegraphics[width=\textwidth]{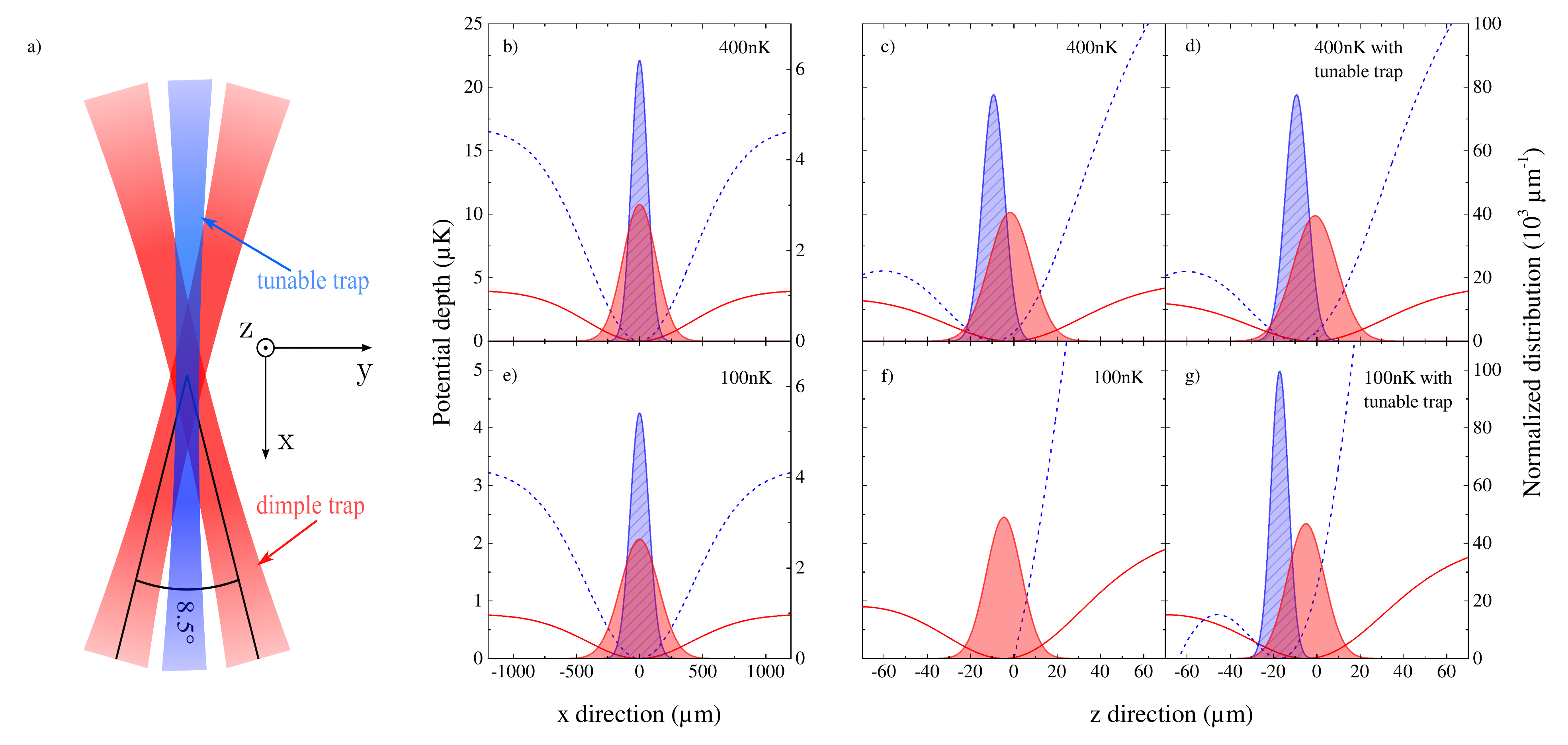}
	\caption{(Color online). a) Top view of the dimple and tunable traps in the horizontal plane (gravity points into the plane). b)-g) Modeled optical dipole plus gravitational potentials for Li (red, solid) and Cs (blue, dashed) atoms at 100~nK and 400~nK and their corresponding atomic density distributions (Li and Cs, shaded red and hatched blue areas, respectively). b) and e) In the horizontal direction the clouds always overlap independent of laser intensity. c) and f) Using solely the dimple trap the spatial overlap is already reduced to 80\% at 400~nK while at 100~nK Cs atoms can no longer be stored due to the influence of the gravitational potential. d) and g) Combined trap. At 100~nK the intensity is 7.0$\times 10^3$~mW/mm$^2$ for the dimple trap and 6.3$\times 10^3$~mW/mm$^2$ for the tunable trap.}
	\label{fig:1}
\end{figure*}

To prepare the Li-Cs mixture we use standard laser cooling techniques in an experimental approach which is similar to the one previously described in Refs. \cite{Repp2013,Pires2014} but extend it with a species-selective laser to compensate the gravitational sag relevant at low temperatures. After the first cooling stage Cs atoms are loaded into a crossed optical dipole trap (reservoir trap) with an 1/$e^2$ waist of 300 $\mu$m, a wavelength of 1064~nm, and a crossing angle of 90$^\circ$. The majority of Cs atoms are optically pumped to the energetically lowest $\left|F=3, m_F=3\right\rangle$ state during degenerate Raman sideband cooling \cite{Treutlein2001}. The Li atoms, populating the two energetically lowest spin states $\left|F=1/2, m_F=1/2\right\rangle$ and $\left|F=1/2, m_F=-1/2\right\rangle$, are loaded into another crossed optical dipole trap, which we refer to as the dimple trap. It has a wavelength of 1070~nm, 1/$e^2$ waist of 62~$\mu$m, crossing angle of $8.5^\circ$, and is located approximately 1~mm away from the reservoir trap. A single spin state is selected by removing the other state through a resonant light pulse. After a separate forced evaporative cooling phase the two optical potentials are combined with a piezo driven mirror and the reservoir trap is slowly switched off. At the end of the final evaporation step $1.6\times 10^4$ ($4\times 10^4$) Cs (Li) atoms are left in the dimple trap at a temperature of 400~nK. The optical dipole potential and the corresponding atomic density distributions of the Li and Cs atoms in x- and z-direction are illustrated in Fig. \ref{fig:1}b and c, respectively.

Further cooling in the same trap configuration is accompanied by a segregation in the z-direction of the two atomic species, especially with a large mass imbalance ($m_{\mathrm{Cs}}/m_{\mathrm{Li}}~\approx 22$), since the influence of gravity on the trapping potential becomes stronger. For a setup as described above the spatial overlap of the two clouds is reduced to $\sim $5\% at an intensity of $\sim$16$\times 10^3$ mW/mm$^2$ and the total potential depth for Li is by factor of $\sim$20 larger than for Cs. For even lower intensities the potential will no longer support the Cs atoms, as shown in Fig. \ref{fig:1}f. These effects restrict the temperatures to about 200~nK for both clouds in order to maintain a substantial overlap.

We circumvent the effect of the gravitational sag for the Cs atoms with a species-selective optical potential which is tuned close to the $\textit{D}_{\mathrm{1}}$ resonance for Cs at 894~nm \cite{LeBlanc2007}. Then the induced dipole moment, and hence the polarizability, for the Cs atoms is much larger than for the Li atoms. Fig. \ref{fig:1}a depicts the beam configuration with an 1/$e^2$ waist of 62 $\mu$m, which is set up along the weak axis of the dimple trap. The wavelength of 921.2~nm is provided by a Ti:sapphire laser and was experimentally chosen for maximizing the difference in the potential depth $U$ of Cs and Li but keeping at the same time the sum of the heating rates of both species as low as possible \cite{LeBlanc2007}. 
 
At this wavelength the atomic polarizability of Cs and Li are $\alpha\approx 4059 $~a.u. and $\alpha\approx 345 $~a.u., respectively. Due to the proportionality $U \propto\alpha$ the trap depth $U$ for Cs is hence by a factor of 12 deeper than for Li \cite{Grimm2000b}. For an intensity of 6.3$\times 10^3$~mW/mm$^2$ the heating rate for Cs and Li is 7~nK/s and 1~nK/s, thus about a factor of 30 and 3 larger compared to the dimple trap, respectively.  
We have nearly independent control over the positions of the two atomic clouds as the position of the Li atoms is mainly determined by the dimple trap and the spatial overlap is retrieved by shifting the tunable laser trap above the dimple trap.

After the mixing of both species is guaranteed with this configuration, the optical power of the dimple trap laser can be reduced to reach lower temperatures around 100~nK for Li. At this power level, Cs atoms would no longer be trapped (see Fig. \ref{fig:1}f), and only in combination with the tunable laser stable storage of the Cs atoms can be provided (see Fig. \ref{fig:1}g). From our estimations the separation between the cloud centers is about 16 $\pm$ 5 $\mu $m. The resulting spatial overlap is then about 45 \% and taken as a constant for the applied magnetic field range. Finally, we obtain $1\times10^4$ ($7\times10^3$) Cs (Li) atoms at a temperature of approximately 100~nK. For the experimental determination of the temperature the widths are recorded for different time-of-flights.

\section{Three-body recombination loss at finite temperature}
\label{sec:2}

In this section, we present our experimental results for the Efimov resonances at finite temperatures through a measurement of the three-body event rate constant close to 843~G. In Sec.~\ref{subsec:exp} we describe the experimental procedure and extraction of the event rate constant $L_3$. In Sec.~\ref{subsec:model}, the universal zero-range theory for the $^{6}$Li-$^{133}$Cs mixture including finite temperatures is introduced \cite{Petrov2015}, with which the data is compared and analyzed in Sec.~\ref{subsec:analysis}. 

\subsection{Experiment}
\label{subsec:exp}
The experimental sequence starts with a cold mixture of Li atoms in state $\left|F=1/2, m_F=1/2\right\rangle$ and Cs atoms at roughly 847~G according to Sec.~\ref{sec:1}. At this magnetic field the trap depth is increased within 150~ms to stop ongoing evaporation as well as to allow for the stabilization of residual magnetic field fluctuations. Here, the temperature is about 120~nK for each cloud and the secular trap frequencies for Cs and Li in (x,y,z) direction are (5.7, 115, 85)~Hz and (25, 160, 180)~Hz, respectively (note that the frequency in the y-direction for Cs has not been measured but has been derived from the model of the optical potentials reproducing closest the five other experimental trapping frequencies). Given that locally, $T / T_{\mathrm{F}} \gtrsim 2$ for Li and $T / T_{\mathrm{C}} \gtrsim 3$ for Cs,  we will consider that the mixture is in the non-degenerate regime.
After a fast ramp to the final magnetic field where the atoms are stored for variable hold times of up to 300~ms we image the atomic clouds on two cameras and deduce the atom number of each species.

\begin{figure}
\centering
\includegraphics[width=\linewidth]{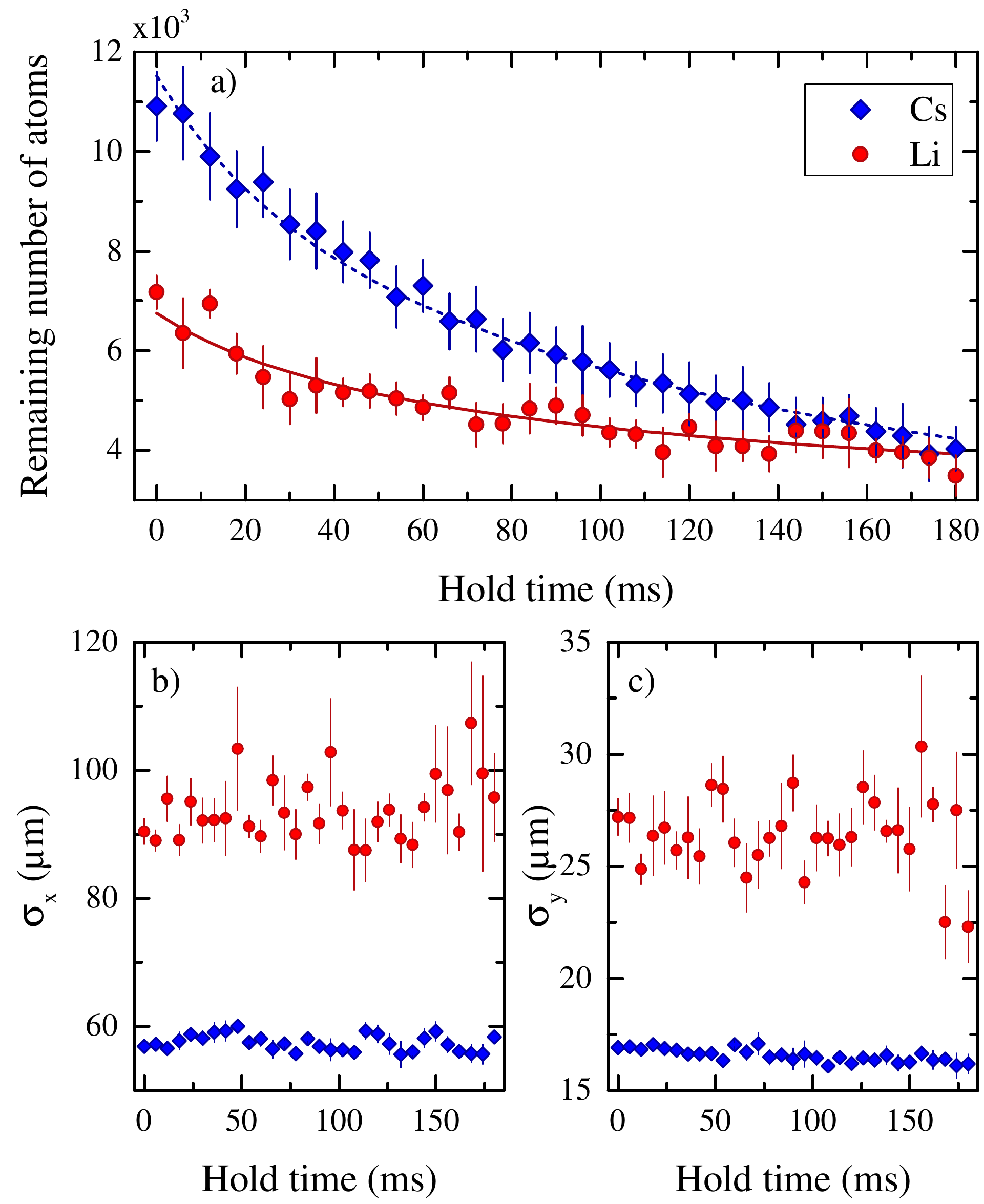}

\caption{(Color online). Temporal evolution of a Li-Cs mixture at 843.049~G and 120 nK. a) Red solid (Li) and blue dashed (Cs) lines show the result of a simultaneous fit of the coupled rate equations (\ref{lossrates1},\ref{lossrates2}) to the atom number of Li (red circles) and Cs (blue diamonds). 
b) and c) Respective $1/e$-widths of the Li and Cs atom clouds in two directions.  
Each data point gives the mean of at least three independent measurements and the error bars represent one standard error. Symbols and color coding are the same as in a).}
\label{fig:2}
\end{figure}

 We measure that the ratio of lost Cs and Li atoms is 2:1, as shown in Fig. \ref{fig:2}. This fact indicates that the loss process of one Li and two Cs atoms dominates. In order to quantify three-body recombination we numerically fit the loss curves for both atomic species simultaneously according to the coupled rate equations
\begin{eqnarray}
\dot{N}_{\mathrm{Li}} &=& -\mathcal{L}_{1}^{\mathrm{Li}}{N}_{\mathrm{Li}} -\mathcal{L}_3 {N}_{\mathrm{Li}} {N}_{\mathrm{Cs}}^{2}
\label{lossrates1}
\\
\dot{N}_{\mathrm{Cs}} &=& -\mathcal{L}_{1}^{\mathrm{Cs}}{N}_{\mathrm{Cs}} - 2\mathcal{L}_3 {N}_{\mathrm{Li}} {N}_{\mathrm{Cs}}^{2} - \mathcal{L}_3^{\mathrm{Cs}}{N}_{\mathrm{Cs}}^{3}
\label{lossrates2}
\end{eqnarray}
where $\mathcal{L}_3$ is converted to the event rate constant $L_3^{\mathrm{expt}}$ by integration over the spatial coordinates assuming the atomic distributions of the Li and Cs atoms based on the model for the bichromatic traps presented in Sec. \ref{sec:1}.
We theoretically estimate the effect of the Li-Cs interaction
on the Li density profile to be enhanced due to the large mass ratio, but still smaller than our current experimental uncertainties; we therefore neglect this effect in the present analysis and leave its detailed study for future work.
 The value $\mathcal{L}_{1}^{\mathrm{Li,Cs}}$ corresponds to the one-body loss rate of each species in the trap and $\mathcal{L}_3^{\mathrm{Cs}}$ to the three-body loss rate for a pure Cs sample at the same conditions. These parameters are determined for our experimental configuration in independent measurements, thus the parameter $\mathcal{L}_3$ as well as the initial atom numbers $N_{0,\mathrm{Li}}$ and $N_{0,\mathrm{Cs}}$ are the only free fitting parameters. The inter- and intraspecies two-body losses are ignored here because the atoms are in the energetically lowest spin states and only exhibit elastic two-body collisions. Moreover, since $^6$Li atoms are identical fermions obeying the Pauli exclusion principle, we neglect any local few-atom processes involving two or more of them, such as, for example, the three-body recombination in the Li-Li-Cs system. In order to estimate error bars of $L_3^{\mathrm{expt}}$ we apply standard methods of bootstrapping and use one standard deviation of the re-sampled distribution as the error \cite{Wu1986}.

In addition, we record the $1/e$-widths of the cloud during the atomic loss (see Fig. \ref{fig:2}b and c) after 5~ms and 0.3~ms of time-of-flight for Cs and Li, respectively. An increase of the width would signal heating due to three-body recombination. However, on the maximal timescale in the Li-Cs mixture of 300~ms the widths of both species stay constant within 10~\%, thus resulting in temperature variations of $\sim$20~\% for our experimental parameters, which endorses us to consider the temperature to be time-independent in our analysis. We also note that in the determination of $L_3^{\mathrm{Cs}}$ the width decreases by 20~\% on the timescale of 1500 ms.

 The scattering length is inferred from the magnetic field $B$ via the relation  $a(B) = a_{\mathrm{bg}}[1-  \Delta/(B-B_{\mathrm{FR}})]$. The conversion has been precisely determined via radio-frequency association of the universal dimers around 843~G and a coupled-channels calculations leading to $B_{\mathrm{FR}}=$ 842.829~G, $\Delta = - 58.21$~G, and  $a_{\mathrm{bg}} = -29.4~a_0$ \cite{Ulmanis2015}. The magnetic field stability is 16~mG resulting from long-term magnetic field drifts, residual field curvature along the long axis of the cigar-shaped trap and calibration uncertainties.

\begin{figure}
	\centering
	\includegraphics[width=\linewidth]{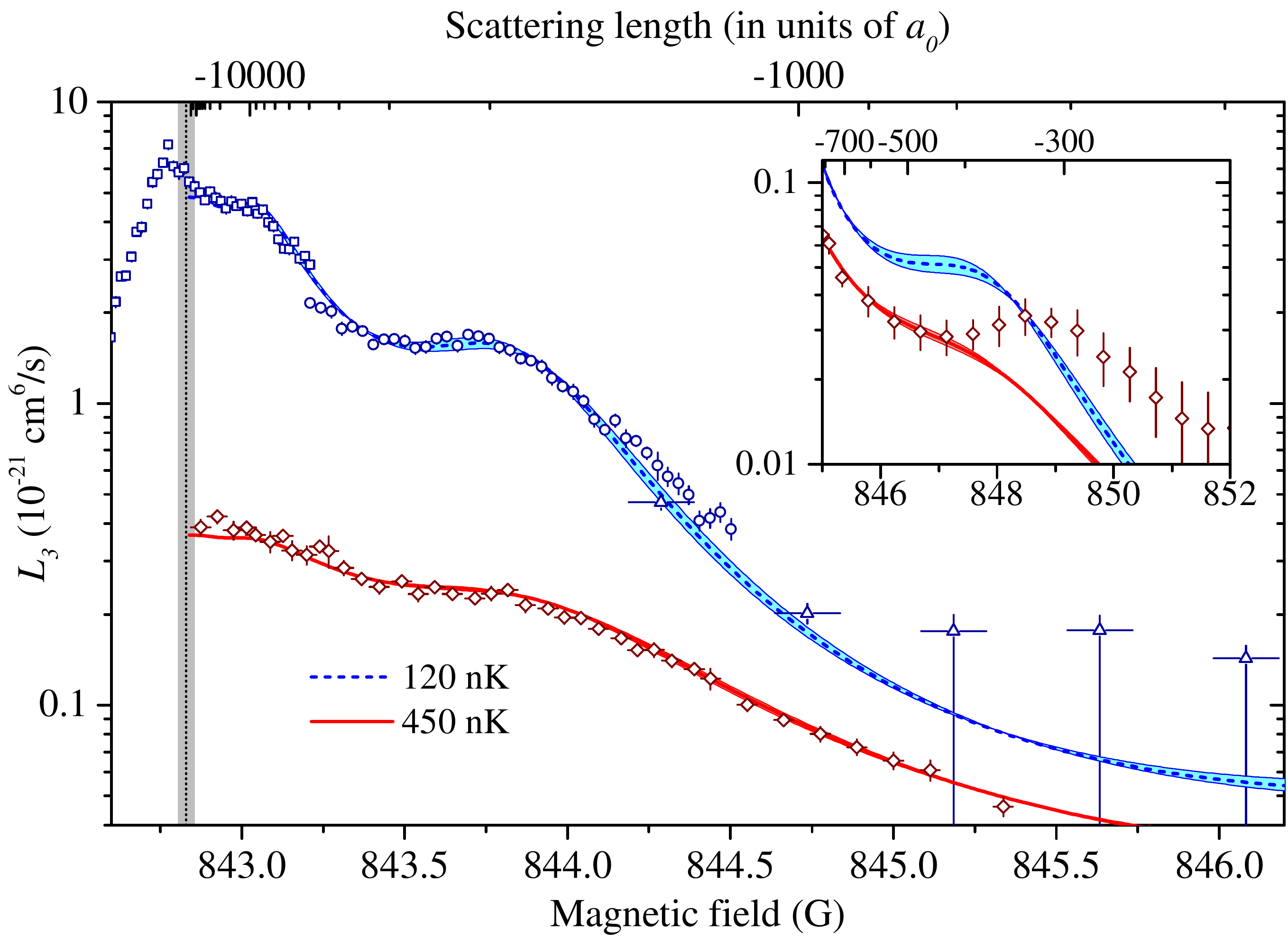}
	\caption{(Color online). Li-Cs-Cs three-body event rate constant $L_3$ at 450~nK (red) and at 120~nK (blue). Experimental data (red diamonds and blue squares, circles, triangles) have been scaled in the vertical direction using a fit to the zero-range theory (solid red lines and dashed blue lines). The shaded region around the fits represent the uncertainty of the Feshbach resonance position. Note that the maximum of $L_3$ does not correspond to the Feshbach resonance position, see Ref. \cite{Ulmanis2015}. The first and second excited Efimov states are clearly visible at $\approx 843.8 $~G and $\approx 843.0 $~G, respectively. Data beyond 845~G represent an upper bound of the loss coefficient at 120~nK due to additional loss dynamics. The inset shows that the Efimov ground state at $\approx 848.9 $~G at 450~nK deviates from the zero-range theory. The 450~nK data are taken from our previous publication \cite{Pires2014}. The pole of the Feshbach resonance is indicated by the dotted line and the uncertainty by the gray shade.}
	\label{fig:3}       
\end{figure}

In Fig.~\ref{fig:3} we show the measured Li-Cs-Cs event rate constants $L_3$ for 450~nK (red) \cite{Pires2014} and 120~nK (blue) as a function of the magnetic field (bottom axis) and the Li-Cs scattering length (top axis). Note that in Ref. \cite{Pires2014}, $L_3$ has been defined as the loss rate constant for the Cs atoms, whereas in this work, $L_3$ is the event rate constant which is twice smaller (factor of two in Eq.~(\ref{lossrates2})). The spectra have one distinct feature in common located at roughly $ 843.8 $~G or $ -$1800~$a_0 $, which corresponds to the first excited Li-Cs-Cs Efimov resonance. For the data taken at 120~nK another feature emerges for an even larger scattering length, now clearly demonstrating the second excited Efimov resonance, which has been anticipated from our previous measurements of loss spectra~\cite{Pires2014}. It is located at about $ 843.0 $~G or $ -$9210~$a_0 $. The Efimov ground state around 848.9~G is only visible in the data set with 450~nK (see inset Fig.\ref{fig:3}). At 120~nK the loss coefficient cannot be reliably extracted from the data below $\sim$~2~$\times10^{-22}$~cm$^6$/s, since other loss processes are dominating, i.e. $\mathcal{L}_{1}^{\mathrm{Cs}}{N}_{\mathrm{Cs}} + \mathcal{L}_3^{\mathrm{Cs}}{N}_{\mathrm{Cs}}^{3} \gg  2\mathcal{L}_3 {N}_{\mathrm{Li}} {N}_{\mathrm{Cs}}^{2} $. In this regime, our fitting procedure (Eqs.~(\ref{lossrates1}) and (\ref{lossrates2})) cannot distinguish the interspecies three-body recombination rate from intraspecies background dynamics. We estimate from our measurements that such a situation realizes once the interspecies loss term acquires a value that is about a factor of 10 smaller than the sum of intraspecies ones. Therefore, the data points at magnetic fields larger than 845~G represent an upper bound of the actual three-body loss rate constant. We note, however, that this restraint is of a purely technical nature and could be at least partially overcome by improving the shot-to-shot stability of the atom numbers, by decreasing ${N}_{\mathrm{Cs}}$ and $\mathcal{L}_{1}^{\mathrm{Cs,Li}}$, or by increasing ${N}_{\mathrm{Li}}$ and the overlap. For the data set at 450~nK the estimated limit is $\sim$~1 $\times10^{-25}$cm$^6$/s, which is well outside the range of our present measurement.

\subsection{Theory}
\label{subsec:model}

We compare the data with the event rate constant derived in the zero-range approximation and based on the $S$-matrix formalism with thermal averaging \cite{Petrov2015}:
\begin{eqnarray}
L_3=  4 \pi^2 \cos^3 \phi \frac{\hbar^7}{\mu^4(k_\mathrm{B}T)^3}(1-e^{-4\eta})\nonumber\\
\times \int\limits_{0}^{\infty} \frac{1-\left |s_{11}\right |^2}{\left| 1+(kR_0)^{-2is_0}e^{-2\eta} s_{11}\right |^2} e^{-\hbar^2k^2/2\mu k_\mathrm{B}T}k dk,
\label{theoeventrate}
\end{eqnarray}
where the angle $\phi$ is defined through the masses by $\sin \phi = m_{\mathrm{Cs}}/(m_{\mathrm{Li}}+m_{\mathrm{Cs}})$, $\mu = m_{\mathrm{Li}}m_{\mathrm{Cs}}/(m_{\mathrm{Li}}+m_{\mathrm{Cs}})$ is the Li-Cs reduced mass, $k_{\rm B}$ is the Boltzmann constant, $R_0$ is the three-body parameter, $\eta$ is the inelasticity parameter, and the quantity $s_0$ is related to the Efimov scaling factor, which in the present case of Li-Cs-Cs with resonant Li-Cs and Cs-Cs interactions equals $\exp (\pi/ s_0)\approx 4.799$. The universal function $s_{11}$ depends on $ka_{\mathrm{LiCs}}$, $ka_{\mathrm{Cs}}$, and the mass ratio. Here $k$ is the three-body collision wave vector defined such that the total energy in the center of mass reference frame equals $\hbar^2k^2/2\mu$. In practice, we use the known magnetic field dependence of $a_{\mathrm{LiCs}}$ \cite{Ulmanis2015} and $a_{\mathrm{Cs}}$ \cite{Berninger2013} and tabulate $s_{11}$ as a function of the magnetic field and $k$ (see Ref.~\cite{Petrov2015} for details on the calculation).

 \subsection{Analysis}
 \label{subsec:analysis}

To analyze the data Eq.~(\ref{theoeventrate}) is used to fit each of the two data sets and extract  $\eta^{(120)} $, $\eta^{(450)} $,  $R_0^{(120)}$, and $R_0^{(450)}$. We also introduce prefactors $\gamma^{(120)}$ and $\gamma^{(450)}$ to rescale the experimental rate~ $ L_3^{\mathrm{expt}}$ to the theoretical one via relation $L_3 = \gamma L_3^{\mathrm{expt}}$. The temperature is a fixed parameter that is given by the experimental conditions, which we determine via independent time-of-flight measurements, and is denoted in the upper indexes of the fitted parameters. Minimization of the error in terms of the reduced $\chi^2 \approx 4.4$ in a weighted fit yields
 $\eta^{(120)} =0.61$, $\eta^{(450)} =0.86$, $R_0^{(120)} =125\, a_0$, and $R_0^{(450)} =130\, a_0$.
For the colder data set we obtain three different constants $\gamma^{(120)}=\{1.9,1.2,0.6\}$ that correspond to measurements taken on different days (see blue squares, circles, and triangles in Fig. \ref{fig:3}), while for the data at higher temperatures we get $\gamma^{(450)}=0.7$. The difference between the fitted prefactors and unity are well within the systematic error of $L_3^{\mathrm{expt}}$, which amounts to a factor of three due to experimental uncertainties of the temperatures, trapping frequencies, overlap and atom numbers.

 The extraction of statistical errors and cross-correlations would require an involved treatment of the experimental uncertainties and error propagation, since the parameters may be connected in a nontrivial way. Instead, we repeat the fit for two different Feshbach resonance pole positions, 842.829 $\pm$ 0.023~G, which correspond to the error for the determination of the pole position \cite{Ulmanis2015}. This changes all fit parameters by less than 2~\%. 
Also, an exclusion of the values above 847~G and 450~nK (red diamonds), where deviations for the Efimov ground state become prominent, modify the fitted values by less than 2\% and improve the reduced $\chi^2$ to only $\approx 4.2$. In a second estimate, we varied all parameters individually such that the reduced $\chi^2$ is worsened by 50\%. The fit parameters change at most by about 32\%.

As shown in Fig. \ref{fig:3} we can now compare in a heuristic way the experimentally determined event rate with the model.  It recaptures not only the full temperature dependence but also the most of the series of an Efimov spectrum with mass imbalance. Good agreement is found for both experimental data sets including the first and second excited Efimov states, while the Efimov ground state cannot be covered (see inset). This behavior is a strong indication that the excited Efimov resonances follow the universal progression, while the ground state is potentially modified by non-universal effects. This would not be very surprising, given that at the ground-state resonance $|a_{\mathrm{LiCs}}|$ is only 3.5 times larger than the Cs-Cs van~der~Waals range, and that the present Li-Cs Feshbach resonance is intermediate between entrance-channel and closed-channel dominated~\cite{Tung2013}. In homonuclear systems, small deviations from the universal zero-range theory have been observed for a ground state Efimov triatomic resonance~\cite{Huang2015a} where $|a|$ is about 10 times the van~der~Waals range~\cite{Berninger2011,Wang2012}, while large deviations were reported for atom-dimer Efimov resonances, see~\cite{Knoop2009,Machtey2012,Machtey2012a,Zenesini2014} and Refs. therein.

In both of our data sets $\eta$ is significantly larger compared to values obtained in previously studied systems, e.g. \cite{Kraemer2006,Wenz2009,Gross2009,Barontini2009,Pollack2009,Berninger2011,Wild2012,Roy2013,Williams2009,Zaccanti2009,Gross2010,Bloom2013,Huang2014,Eismann2015}. This could indicate efficient loss channels into deeply bound molecules. Alternatively, a large $\eta$ in the fits might also be caused by an overlapping tetramer or higher order recombination resonance \cite{Blume2014} resulting in an effective increase in the three-body event rate. However, with the current resolution of the experimental data and the achievable temperature ranges, these resonances cannot be distinguished.

A frequently discussed property of the Efimov resonances is the distance between them. The peaks are expected to be equidistant on a log-scale if one changes all scattering lengths proportionally to each other. In this case the scaling factor for the Li-Cs-Cs system is approximately $4.80$. A slightly larger scaling factor, $\approx 4.88$, is expected in the case of non-resonant Cs-Cs interaction \cite{Efimov1973,Yamashita2013,Petrov2015}. In order to quantify positions of the Efimov resonances we first fit the data with Eq.~\eqref{theoeventrate} thus determining $R_0$. Then the series of positions corresponding to the Efimov resonances are found by simultaneously setting in the theory $\eta \rightarrow 0$ and $T \rightarrow 0$. The results for $R_0 = 125\, a_0$ are listed in Table \ref{tab:1}. A three-body parameter of $R_0 = 130\, a_0 $ changes these values by about 1~\%.

\begin{table}
	\caption{\label{tab:1}
		Li-Cs-Cs Efimov resonance positions in terms of the calculated magnetic field $B_{\mathrm{zr}}^{(n)}$, scattering length $ a^{(n)}_- $ , and ratio $ \lambda_{\mathrm{zr}}^{(n)} $ from the zero-range theory, where $R_0$ and $\eta$ are fixed from the fit to the experimental data. Also given are the respective scattering lengths of Cs $ a_{\mathrm{Cs}}$ and the positions of Efimov resonances through Gaussian fits $B^{(n)}_{\mathrm{G}, 120~\mathrm{nK}}$ to experimental data at 120~nK~\cite{Pires2014,Ulmanis2015}. The quantities in the parenthesis always represent the statistical and systematical error arising from the determination of the Efimov resonance positions and magnetic field uncertainty. }
	\begin{ruledtabular}
		\begin{tabular}{llllll}
			$ n $ & $B_{\mathrm{zr}}^{(n)}$ (G) & $ a^{(n)}_- $ $ (a_0) $ & $ \lambda_{\mathrm{zr}}^{(n)} $ & $ a_{\mathrm{Cs}} (a_0) $ & $B_{\mathrm{G}, 120~\mathrm{nK}}^{(n)}$(G) \\ \hline
			0     & 848.167                     & -350                    &                                 & -1218                     & -                         \\
			1     & 843.808                     & -1777                   & 5.08                            & -1493                     & 843.772(10)(16)           \\
			2     & 843.015                     & -9210                   & 5.18                            & -1548                     & 843.040(10)(16)           \\
			3     & 842.866                     & -46635                  & 5.06                            & -1559                     & -
		\end{tabular} 
	\end{ruledtabular}
\end{table}

The Li-Cs and Cs-Cs scattering lengths do not change proportionally, and the ratio $ a^{(n)}_-/a^{(n)}_{\rm Cs} $ is not a constant. Therefore, according to the universal zero-range theory the quantity $\lambda_{\mathrm{zr}}^{(n)}\equiv a_{-}^{(n)}/a_{-}^{(n-1)}$ does depend on $n$. Subsequent ratios with higher $n \ge 4$ gradually tend to 4.88, as expected for the Li-Cs-Cs system with resonant interspecies and finite intraspecies interactions. This behavior is expected on approach to the Li-Cs Feshbach resonance pole, where $a_{\mathrm{Cs}}$ tends to a constant and becomes negligible compared to the diverging $a_{\mathrm{LiCs}}$. The manifestation of scaling factor 4.80 would require much larger values of $a_{\mathrm{ Cs}}$ or the ability to change both $a_{\mathrm{LiCs}}$ and $a_{ \mathrm{Cs}}$ proportionally.

Due to the lack of a finite-temperature model and an insufficient agreement with the analytical zero-temperature model, which was assuming $a_{\mathrm{Cs}}=0$ \cite{Helfrich2010}, Gaussian fit profiles were used previously to determine the Efimov resonance positions $B^{(n)}_{\mathrm{G}}$ ~\cite{Pires2014,Tung2014}. 
Following this approach for the new data at 120~nK we obtain the values as given in Table~\ref{tab:1}. In comparison to the values that were extracted from the presently implemented zero-range model, the fitting of Gaussian profiles seems to be adequate for describing resonance positions of higher excited states within the current experimental uncertainties. 
In contrast, the Efimov ground state is not described by the zero-range model, therefore it is not surprising that its position  $B^{(0)}_{\mathrm{G}, 450~\mathrm{nK}}=848.90(6)(3)$ G ~\cite{Ulmanis2015}, as determined by a Gaussian fit, also deviates from the theoretical one $B_\mathrm{zr}^{(0)}$. 
This hint at non-universality has been quantified previously by a larger experimental ratio $a_{-}^{(1)}/a_{-}^{(0)}=5.5(2)$~\cite{Ulmanis2015}.

\section{Conclusion}
\label{sec:3}

In summary, we have measured  three-body recombination spectra in an ultracold mixture of Li and Cs atoms at $T=120$~nK, which is a factor of four lower than in our previous work \cite{Pires2014}. The lower temperature was achieved by compensating the gravitational sag via species-selective optical potentials, thus enabling us to attain temperatures at a 100~nK level, while retaining partial overlap of the atom clouds. This procedure allowed us to undoubtedly reveal the second excited Efimov resonance.  The new data in combination with the previous results from \cite{Pires2014} were compared with the universal zero-range theory~\cite{Petrov2015}. 
The excellent agreement close to the pole of an interspecies Feshbach resonance demonstrates that the model captures the essential features of the three-body recombination  in the universal regime. A possible alternative to the currently employed approach is the optical model with phenomenological imaginary potentials~\cite{Mikkelsen2015}.

At small scattering lengths we observed clear deviations of the Efimov ground state from the universal zero-range theory, as already anticipated in our earlier measurements~\cite{Pires2014,Ulmanis2015}. The deviations are likely due to finite-range effects, which calls for further studies taking into account the van der Waals tail of the interactions~\cite{Wang2012,Wang2012d,Naidon2014a},
the multi-channel nature of the employed Feshbach resonance~\cite{Lee2007,Jona-Lasinio2010,Schmidt2012,Sorensen2013a},  both of these features~\cite{Wang2014a}, or effective-range corrections~\cite{Ji2010,Ji2015,Kievsky2015}. 

In contrast to previously experimentally studied systems that were using single-species gases or K-Rb mixtures, the inelasticity parameter $\eta^{(120)}=0.61$ or $\eta^{(450)}=0.86$ is significantly larger for the Li-Cs mixture. The origin of this behavior is still an open question. Possible explanations may include the large Cs-Li mass ratio or higher order processes.
Further investigations are desirable to understand how these findings change when $a_{\mathrm{Cs}}$ becomes zero or even positive.

\section*{Acknowledgments}
We are grateful to C. Greene, R.~Grimm, Y. Wang, and S. Whitlock for fruitful discussions.
J. U. acknowledges support by the DAAD, D.S.P. and F.W. acknowledge support by the IFRAF Institute, S. H. and  R. P. by the IMPRS-QD, E. K. is indebted to the Baden-W\"urttemberg Stiftung for the financial support of this research project by the Eliteprogramme for Postdocs.
This work is supported in part by the Heidelberg Center for Quantum Dynamics. 
The research leading to these results has received funding from the European Research Council under European Community's Seventh Framework Programme (FR7/2007-2013 Grant Agreements no.341197 and 307551).


\bibliography{Bib_EfimovZeroRange}


\end{document}